\documentclass[useAMS,usenatbib,usegraphicx,letter]{mn2e}
\usepackage{amssymb}
\newcommand\Alfven{Alfv\'en }
\newcommand\Alfvenic{Alfv\'enic }

\title[A prescription for turbulent heating]{A prescription for the turbulent heating of astrophysical plasmas}
\author[G. G. Howes]{G. G. Howes
\thanks{E-mail: gregory-howes@uiowa.edu} 
\\
Department of Physics and Astronomy, University of Iowa, 505 Van Allen Hall, Iowa City, IA 52245, USA}
\begin{document}


\pagerange{\pageref{firstpage}--\pageref{lastpage}} \pubyear{2009}

\maketitle

\label{firstpage}

\begin{abstract}
The ratio of ion to electron heating due to the dissipation of
Alfv\'enic turbulence in astrophysical plasmas is calculated based on
a cascade model for turbulence in weakly collisional
plasmas. Conditions for validity of this model are discussed, a
prescription for the turbulent heating is presented, and it is  applied
to predict turbulent heating in accretion disks and the interstellar
medium.
\end{abstract}


\section{Introduction}

Plasma is a ubiquitous form of matter in the universe, nearly always
found to be magnetized and turbulent.  In a wide variety of space and
astrophysical plasmas, the dissipation of turbulent fluctuations makes
a significant contribution to the thermodynamic heating of the plasma,
requiring an understanding of this behavior to interpret a large body
of astronomical observations. Recently, there has been much vigorous
activity in the study of turbulent heating in a wide variety of systems,
 including galaxy clusters
\citep{Chandran:2004,Dennis:2005}, the interstellar medium
\citep{Minter:1997,Struck:1999,Spangler:2003,Scalo:2004,Lerche:2007},
accretion disks
\citep{Balbus:1994,Quataert:1998,Gruzinov:1998,Quataert:1999,Sharma:2007,Rossi:2008},
molecular clouds
\citep{Pan:2009,Lemaster:2009}, the solar 
corona \citep{Velli:2003,Verdini:2009,Chandran:2010b}, the solar wind
\citep{Saito:2008,Valentini:2008,Breech:2009,Cranmer:2009,Lehe:2009,Chandran:2010a},
and the magnetospheres of the Earth \citep{Chaston:2008} and of other
planets \citep{Saur:2004}.

For many of these plasma environments, the dissipation of turbulent
fluctuations occurs at length scales, in the direction parallel to the
local mean magnetic field, that are small compared to the particle
mean free paths. The resulting weakly collisional dynamics
then requires a kinetic description
\citep{Howes:2008d,Howes:2008a}, with the dissipation occurring via collisionless
wave-particle interactions
\citep{Howes:2008c,Schekochihin:2009}. These kinetic damping
mechanisms generally transfer energy to the plasma ions and electrons
at different rates. The time scale for the collisional equilibration
of temperature between species is often long, allowing the plasma ions
and electrons to maintain distinct heating rates and temperatures. The radiation
emitted from the hot, magnetized plasma will often depend strongly on
the nature of the turbulent plasma heating, exerting a dominant
influence on the observational signature of the object.

Until very recently, the question of differential heating of the ions
and electrons by plasma turbulence had not been widely considered,
with the exception of a pioneering series of papers by Quataert and
Gruzinov \citep{Quataert:1998,Gruzinov:1998,Quataert:1999}. The need
for an accurate description of the turbulent plasma heating by kinetic
mechanisms motivates the two primary aims of this Letter: (1) to
calculate the relative heating of ions and electrons due to the
kinetic dissipation of plasma turbulence; and (2) to provide a
prescription for turbulent plasma heating that may be used as a
sub-grid-scale model for heating in simulations of astrophysical
plasma turbulence.

\section{Turbulent Heating Model}
The turbulent heating prescription determined in this Letter is based
on the \citet{Howes:2008b} model for the turbulent cascade of energy
in a weakly collisional plasma. Whereas that study focused on
conditions relevant to turbulence in the near-Earth solar wind, here
we generalize to a more broad range of plasma parameters for general
astrophysical applications. A brief description of the model and its
underlying assumptions follows---more detailed discussion is found in
\citet{Howes:2008b}.

The cascade model makes three primary assumptions: (1) the Kolmogorov
hypothesis that the energy cascade is determined by local interactions
\citep{Kolmogorov:1941}; (2) the turbulence maintains a state of
critical balance at all scales  \citep{Goldreich:1995}; and (3) the
linear kinetic damping rates are applicable in the nonlinearly
turbulent plasma. Modern theories of MHD turbulence 
\citep{Goldreich:1995,Boldyrev:2006} predict an  anisotropic cascade of 
turbulent energy through wave vector space in a magnetized plasma
\footnote{Slightly different predictions for inertial range spectral
slopes $k_\perp^{-5/3}$ or $k_\perp^{-3/2}$ are unlikely to yield
significant differences in the turbulent heating ratios predicted by
this model.}.  The resulting state of critical balance is supported by
numerical simulations \citep{Cho:2000} and observations in the solar
wind
\citep{Horbury:2008,Podesta:2009a}.  A recent study of test particle heating 
in MHD turbulence supports the use of linear theory to estimate the
damping and heating in a turbulent plasma \citep{Lehe:2009}.

Astrophysical turbulence typically supports a large inertial range at
perpendicular scales larger than the ion Larmor radius $k_\perp
\rho_i< 1$, leading to turbulent fluctuations at $k_\perp \rho_i
\gtrsim 1$ that are highly anisotropic with $k_\parallel \ll k_\perp$. Such 
anisotropic fluctuations are optimally described by a low-frequency
expansion of kinetic theory called gyrokinetics
\citep{Rutherford:1968,Frieman:1982,Howes:2006,Schekochihin:2009}. In this limit, 
all collisionless dissipation occurs via the Landau resonance, and the
cyclotron resonance is negligible.  Limitations of the application of
gyrokinetic theory in the solar wind are discussed in detail in
\citet{Howes:2008b}.   Energy spectra from the  first nonlinear gyrokinetic 
simulations of \Alfvenic turbulence at the scale of ion Larmor radius
are modeled with striking agreement by the cascade model used in this
Letter \citep{Howes:2008a}. 

 A key simplification of the physics in the
gyrokinetic limit is that the normalized eigenfrequency of the linear,
collisionless gyrokinetic dispersion relation
\citep{Howes:2006} is independent of $k_\parallel$, giving 
$\overline{\omega}(k_\perp, \beta_i, T_i/T_e)=
\omega(k_\perp,k_\parallel, \beta_i, T_i/T_e)/k_\parallel v_A$; the same 
holds for the damping rate, $\overline{\gamma}= \gamma/k_\parallel
v_A$. This enables the development of a one-dimensional model of the
turbulent magnetic field energy (in velocity units), $b_k^2
\equiv \delta B_\perp^2 (k_\perp) /4 \pi n_i m_i$, where $\delta
B_\perp^2(k_\perp)/8 \pi$ is the energy density of the magnetic field
fluctuations perpendicular to the mean field integrated over
$k_\parallel$.

The model assumes a fully ionized plasma of protons and electrons with
Maxwellian equilibrium velocity distributions of temperature $T_s$ (in
energy units). Plasma thermal velocities are taken to be
non-relativistic, $v_{ts}= \sqrt{2T_s/m_s} \ll c$.  At the scales of
dissipation (of order the ion Larmor radius), the mean magnetic field
can be modeled as a straight, uniform field of magnitude $B_0$. The
turbulence is assumed to be sub-Alfv\'enic, driven isotropically at a
wave number $k_0$ with an amplitude satisfying the critical balance
condition, and balanced (equal \Alfven wave energy fluxes in opposite
directions along the magnetic field).  The model follows the magnetic
energy in the Alfv\'enic cascade as it transitions from
\Alfven waves to kinetic \Alfven waves at $k_\perp \rho_i \sim 1$
\citep{Howes:2008a,Howes:2008c,Schekochihin:2009}. The numerical algorithms used to 
calculate the linear gyrokinetic and Vlasov-Maxwell dispersion
relations are described in \citet{Quataert:1998} and
\citet{Howes:2006}.

The continuity equation for the evolution of the magnetic energy
versus perpendicular wavenumber is given by
\begin{equation}
\frac{\partial b_k^2}{\partial t} = 
-k_\perp \frac{\partial \epsilon(k_\perp) }{\partial k_\perp} + S(k_\perp) - 
2{\gamma} b_k^2,
\label{eq:modelcont}
\end{equation}
where the energy cascade rate is $\epsilon(k_\perp) = C_1^{-3/2}
k_\perp \overline{\omega} b_k^3$, the energy injection rate is $S$
(non-zero only at the driving scale $k_\perp=k_0$), and the linear
kinetic damping rate is $\gamma$.  Assuming critical balance at all
scales, the steady state solution for the energy cascade rate is given
by
\begin{equation}
\epsilon(k_\perp) = \epsilon_0 
\exp{\left\{-\int_{k_0}^{k_\perp}2C_1^{3/2}C_2
\frac{\overline{\gamma}(k_\perp ')}{\overline{\omega}(k_\perp ')}
\frac{dk_\perp '}{k_\perp '}\right\}},
\label{eq:modelcont3}
\end{equation}
where $C_1$ and $C_2$ are order unity dimensionless Kolmogorov
constants and $\epsilon_0$ is the rate of energy input at $k_0$.  The
solution of $\epsilon(k_\perp)$ for parameters
$(\beta_i, T_i/T_e)$ is then used, along with the linear gyrokinetic
damping rates due to each species $\overline{\gamma}_s$
\citep{Howes:2006}, to calculate the spectrum of heating by species
$Q_s(k_\perp)=2C_1^{3/2}C_2( \overline{\gamma}_s/\overline{\omega})
\epsilon(k_\perp)/k_\perp$. After integrating over $k_\perp$, we
obtain the total ion-to-electron heating rate due to the kinetic
dissipation of the turbulent cascade, $Q_i/Q_e(\beta_i, T_i/T_e)$, the
primary scientific result of this Letter. Note that
\citet{Quataert:1999} presented a similar model to estimate the
fraction of electron heating for $T_i/T_e=100 $ and $1\le \beta_i \le
10^3$, but did not account for the transition to a kinetic \Alfven
wave cascade at the scale of the ion Larmor radius $k_\perp \rho_i
\sim 1$.

\subsection{Limitations of the Model}
\label{sec:limits}

First, the cascade at weakly collisional scales with $k_\perp \rho_i <
1$ is assumed to consist entirely of Alfv\'enic fluctuations with
\Alfven Mach number less than unity. The energy in compressive modes,
such as the collisionless manifestation of the MHD fast and slow
magnetosonic waves, is neglected based on studies that demonstrate
strong damping of these waves under weakly collisional conditions for
some parameter regimes
\citep{Barnes:1966,Foote:1979}; the applicability of this result over
the entire range of parameters considered here merits further study,
beyond the scope of this work. Here we focus solely on the cascade of
\Alfven waves, which are incompressible to lowest order and remain
undamped down to the scale of the ion Larmor radius $k_\perp \rho_i
\sim 1$ \citep{Schekochihin:2009}. If significant turbulent energy
exists in the compressive wave modes, then its dissipation and the
resulting plasma heating must be accounted for separately;
\citet{Quataert:1998} discusses this matter in more detail. 

Second, the entropy increase required to achieve irreversible
thermodynamic heating can only be accomplished through collisions
\citep{Howes:2006}.  Wave-particle interactions damp the
electromagnetic fluctuations comprising the turbulence and deposit the
associated energy into non-thermal features of the particle
distribution function \citep{Howes:2008c,Schekochihin:2009}.  It is
assumed that the action of collisions on the non-thermal distributions
of a given species lead to heating of that plasma species.

Third, we do not explicitly account for the kinetic energy of the
Alfv\'enic fluctuations at $k_\perp \rho_i < 1$. The restoring force of 
\Alfven waves is magnetic tension, so the energy in an \Alfven wave is 
transferred back and forth between magnetic energy and kinetic energy
at the wave frequency; for an ensemble of many \Alfven waves, the
turbulent fluctuation energy is shared equally between magnetic and
kinetic forms. As the turbulence transitions to scales $k_\perp \rho_i
\gtrsim 1$, the ions decouple from the turbulence, causing the kinetic
energy to become subdominant. It is assumed that the rapid transfer
between kinetic and magnetic energy at the \Alfven wave frequency
enables both the kinetic and magnetic energy at scales $k_\perp \rho_i
< 1$ to be channeled into the dominantly magnetic fluctuations in the
kinetic \Alfven wave regime $k_\perp \rho_i \gtrsim 1$.   Nonlinear
simulations of kinetic turbulence will provide a crucial tool to test
this assumption.

\section{Results}

\begin{figure}
\resizebox{84mm}{!}{\includegraphics*[0.25in,2.in][8.05in,8.1in]{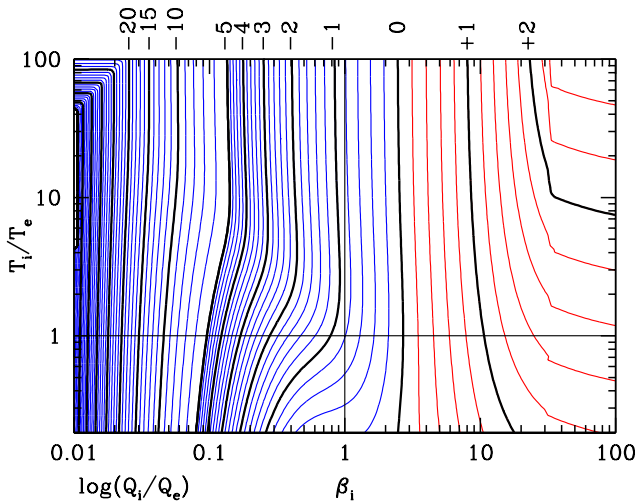}}
 \caption{Contour plot of $\log(Q_i/Q_e)$ over the plane $(\beta_i, T_i/T_e)$.
\label{fig:qiqe} }
\end{figure}

The key scientific result of this Letter, the calculation of ion-to-electron heating due to the kinetic dissipation of the Alfv\'enic
turbulent cascade, $Q_i/Q_e(\beta_i, T_i/T_e)$, is presented in
Fig.~\ref{fig:qiqe} as a logarithmic contour plot over the parameter
range $0.01 \le \beta_i \le 100$ and $0.2 \le T_i/T_e \le 100$. The
values of the Kolmogorov constants used in this calculation,
$C_1=1.96$ and $C_2=1.09$, are derived from a fit of the cascade model
predictions to nonlinear simulations of the transition from the
\Alfven to the kinetic \Alfven wave cascade \citep{Howes:2008a}.

A salient feature of this result is that the heating rate $Q_i/Q_e$ is
primarily a monotonic function of $\beta_i$ and is nearly independent
of $T_i/T_e$. \citet{Quataert:1998} and \citet{Quataert:1999} provide
an in-depth discussion of various kinetic damping mechanisms and their
dependence on the plasma parameters; here we present only the key
concepts necessary to explain this behavior. The ion damping peaks at
$k_\perp \rho_i \sim 1$, while the electron damping peaks at $k_\perp
\rho_i \gg 1$ unless $T_i/T_e \lesssim m_e/m_i$, a condition 
unlikely to be relevant to astrophysical plasmas.  Therefore, any
energy that passes through the peak of the ion damping at $k_\perp
\rho_i \sim 1$ will lead to electron heating.  The damping onto ions
via the Landau resonance is dominated by transit-time damping
involving the interaction of the ion's magnetic moment with the
longitudinal magnetic field perturbation $\delta B_\parallel$ (the
magnetic analog of Landau damping by the parallel electric field)
\citep{Barnes:1966,Quataert:1998}. As the  magnetic moment of the ions
increases with increasing $\beta_i$, so does the strength of the
transit-time damping, leading to its dominant role in controlling the
ion-to-electron heating $Q_i/Q_e$. The value of $T_i/T_e$ controls the
scale at which the electron damping occurs, but any energy passing
through the range of ion damping will ultimately heat the
electrons, leading to little effect on the ratio of total damping
$Q_i/Q_e$.

For values of $T_i/T_e < 0.2$, the slow magnetosonic mode, often
called the ion acoustic wave in this limit, is no longer heavily
damped. In this case, the energy in this compressive mode may no
longer be neglected. Further complicating matters is the fact that the
slow wave and \Alfven wave may nonlinearly exchange energy at the
scale of the ion Larmor radius \citep{Schekochihin:2009}, so the ratio
of ion-to-electron heating may no longer be accurately estimated using
the present model. Nonlinear kinetic simulations of the turbulence at
the scale of the ion Larmor radius will yield valuable guidance to
determine the turbulent heating ratio $Q_i/Q_e$ in this regime.

\begin{figure}
\resizebox{84mm}{!}{\includegraphics*[0.25in,2.in][8.05in,7.7in]{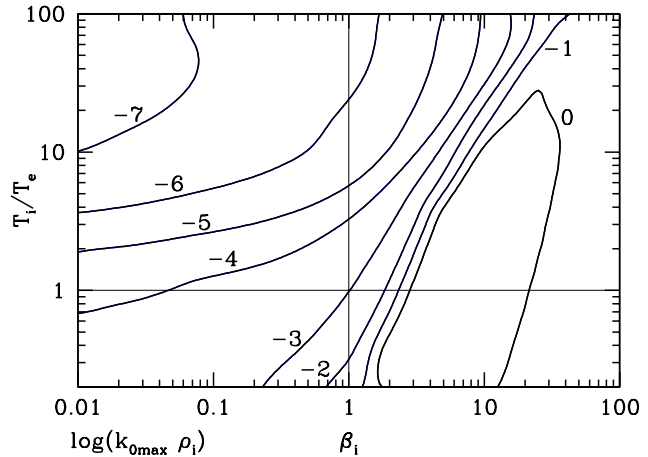}}
 \caption{Contour plot of the $\log(k_{0max} \rho_i)$ over the plane 
$(\beta_i, T_i/T_e)$, a necessary condition for the validity of the heating results 
presented here.
\label{fig:k0} }
\end{figure}

For the gyrokinetic treatment of $Q_i/Q_e$ presented here to be
applicable, the turbulence must be dissipated kinetically (via the
Landau resonance) before the frequencies rise to the point that the
cyclotron resonance alters the dynamics. Since $\omega \propto
k_\parallel$ in the gyrokinetic limit, this constraint may be cast as
a limitation on the maximum isotropic driving wavenumber $k_{0max}$
\citep{Howes:2008b}: for turbulence driven at wavenumbers below
$k_{0max}$, frequencies remain sufficiently low that the cyclotron
resonance never contributes and the gyrokinetic treatment is
valid. The value of $k_{0max}$ is determined using the following
procedure: (1) using the critical balance condition to give
$k_\parallel = k_0^{1/3} k_\perp^{2/3}
\overline{\omega}^{-1/3} (\epsilon/\epsilon_0)^{1/3}$, the maximum
value of $k_\parallel/k_0$ is determined from the steady-state solution
$\epsilon(k_\perp)$; (2) from this point (with fixed $k_\perp$,
$\beta_i$, and $T_i/T_e$), the eigenfrequency using the linear
Vlasov-Maxwell dispersion relation is solved as $k_\parallel$ is
increased until cyclotron resonant effects lead to $[(\omega_{vm} -
\omega_{gk})^2/(\omega_{vm})^2 + (\gamma_{vm} -
\gamma_{gk})^2/(\gamma_{vm})^2 ]^{1/2}> 0.5$; (3) the value of the
parallel wavenumber $k_{\parallel max}$ at this point is used to
compute the corresponding $k_{0max}$. A logarithmic contour plot of
the value of the normalized maximum isotropic driving wavenumber
$k_{0max} \rho_i$ over the $(\beta_i, T_i/T_e)$ plane, presented in
Fig.~\ref{fig:k0}, enables verification of the validity of the
heating results for a turbulent astrophysical plasma with given values
of $\beta_i$, $T_i/T_e$, and $k_{0}$.

\subsection{Prescription for Turbulent Heating}
\begin{figure}
\resizebox{84mm}{!}{\includegraphics*{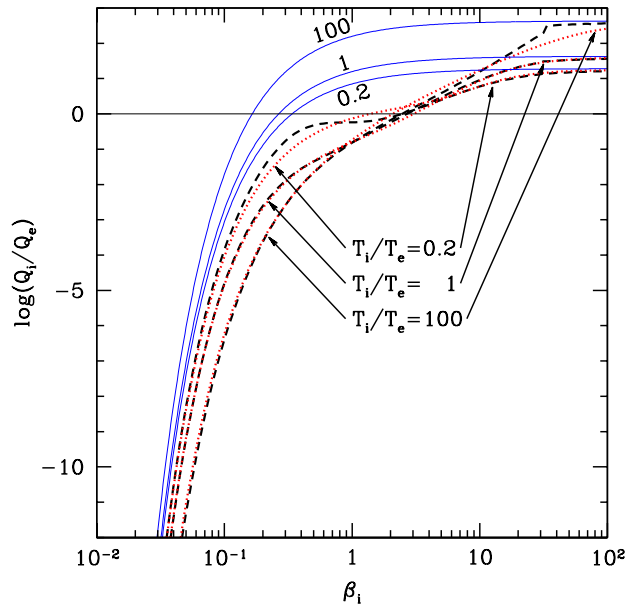}}
 \caption{Calculations of $\log(Q_i/Q_e)$ vs.~$\beta_i$ from the
 cascade model (dashed), the fitted heating prescription
 eq.~(\ref{eq:fit}) (dotted), and eq.~(5) of \citet{Quataert:1998}
 (solid). \label{fig:fit} }
\end{figure}

The second aim of this paper is to provide a prescription for the
turbulent heating $Q_i/Q_e$ presented in Fig.~\ref{fig:qiqe} that
may be used as a sub-grid-scale model for heating in simulations of
astrophysical plasma turbulence. The heating prescription, found by
fitting the results for $Q_i/Q_e$ in Fig.~\ref{fig:qiqe}, is given
by
\begin{equation}
Q_i/Q_e = c_1\frac{c_2^2 + \beta_i^p}{c_3^2 + \beta_i^p} \sqrt{\frac{m_i T_i}{m_e T_e}} e^{-1/\beta_i}
\label{eq:fit}
\end{equation}
where $c_1=0.92$, $c_2=1.6/(T_i/T_e)$, $c_3=18+5 \log (T_i/T_e)$, and
$p=2-0.2 \log (T_i/T_e)$.  A slightly better fit for $T_i/T_e <1$
occurs using the coefficients $c_2=1.2/(T_i/T_e)$ and $c_3=18$. It is
interesting to compare this result to eq.~(5) of
\citet{Quataert:1998}, $Q_i/Q_e = [(m_i T_i)/(m_e T_e)]^{1/2} e^{-1/\beta_i}$,
which gives the predicted value of ion-to-electron heating by transit
time damping in the limit $k_\perp \rho_i <1$. In Fig.~\ref{fig:fit},
$\log(Q_i/Q_e)$ is plotted for $T_i/T_e= 0.2,1,100$ for the three
methods: the cascade model calculations (dashed); the heating
prescription, eq.~(\ref{eq:fit}) (dotted); and eq.~(5) of
\citet{Quataert:1998} (solid). One can see that, although the 
estimation by \citet{Quataert:1998} generally reproduces the behavior
of $Q_i/Q_e$, significant deviations at $\beta_i \sim 1$ are predicted
by the cascade model calculations, and the numerical results are much more
accurately described by the heating prescription given by
eq.~(\ref{eq:fit}).

\section{Discussion}
We consider here the application of eq.~(\ref{eq:fit}) to studies of
astrophysical accretion disks and to the interstellar medium.

Recent two-fluid studies of the magnetorotational instability (MRI) in
accretion disks using shearing box simulations by \citet{Sharma:2007}
have shown that the development of pressure anisotropies, and their
saturation via kinetic instabilities, leads to substantial viscous
heating, representing a direct conversion of kinetic energy due to
differential rotation into heat at large scales. About 50\% of the
heating in the simulations occurred via this mechanism, with the ratio
of ion-to-electron viscous heating given approximately by
$Q_{V,i}/Q_{V,e} \approx 3 (T_i/T_e)^{1/2}$.  Numerical grid-scale
dissipation accounted for the remaining 50\% of the turbulent energy
loss; this loss was not added to the plasma heating because it was
unclear which species should receive this energy.  By applying the
predicted viscous electron heating to Sgr A* using a 1D model for the
radiative efficiency, the observed luminosity of $L\approx
10^{36}$~ergs~s$^{-1}$ implied an accretion rate of $\dot{M}
\sim 10^{-7}$ to $10^{-8} \ M_\odot$~yr$^{-1}$, well below the
estimated Bondi accretion rate \citep{Sharma:2007}.

The 50\% of the energy lost numerically in their simulations
represents the energy transferred to small scales via turbulence,
eventually to be damped by collisionless dissipation mechanisms.
Here we estimate the partitioning of this heating by species using the
turbulent heating model in eq.~(\ref{eq:fit}). The MRI drives
turbulence on the scale height of the disk \citep{Quataert:1998}, so we
take the isotropic driving wavenumber to be $k_0 \sim 1/H \sim 10
^{-8}$~km for Sgr A*.  For an ion temperature of $T_i \sim 10^{12}$~K,
and a magnetic field strength of $B \sim 30$~G, the ion Larmor radius
is approximately $\rho_i \sim 0.4$~km \citep{Schekochihin:2009}.  Following 
\citet{Sharma:2007}, we take the plasma $\beta_i \sim 10$ and $T_i/T_e \sim 10$, so 
for a value of $k_0\rho_i \sim 10{^{-7}}$, we see from
Fig.~\ref{fig:k0} that we are well below the maximum value $k_{0max}
\rho_i \sim 1 $ required for the validity of eq.~(\ref{eq:fit}).

\begin{figure}
\resizebox{84mm}{!}{\includegraphics*[0.25in,2.1in][8.05in,6.in]{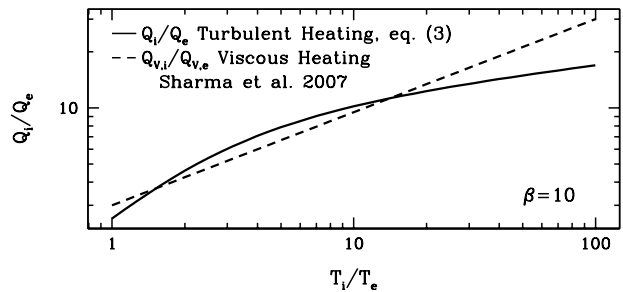}}
\caption{For a $\beta=10$ plasma, the ion-to-electron heating ratio $Q_i/Q_e$ predicted by  eq~(\ref{eq:fit}) (solid) and viscous
heating $Q_{V,i}/Q_{V,e}$ arising from anisotropic pressure due to the
background shear flow (dashed) from
\citet{Sharma:2007}. \label{fig:ad} }
\end{figure}

In Fig.~\ref{fig:ad}, we compare the turbulent heating ratio $Q_i/Q_e$
from eq.~(\ref{eq:fit}) for $\beta=\beta_i (1 + T_e/T_i)=10$ 
to the viscous heating $Q_{V,i}/Q_{V,e}$ from \citet{Sharma:2007}.
The quite surprising result is that, for a $\beta=10$ plasma, the two
distinct mechanisms yield a heating ratio that agrees within a factor
of two over the entire range of $T_i/T_e$.  Therefore, adding in 
the plasma heating due to the damping of the turbulence via kinetic
mechanisms does not change the conclusions of \citet{Sharma:2007}
regarding Sgr~A* discussed above.

Note that, in addition to the turbulent heating prescription presented
here, a complete sub-grid model for turbulence would also require
prescriptions for the sub-grid Reynolds and Maxwell stresses. The
relative insensitivity of the measured stresses to the numerical
resolution in the simulations of \citet{Sharma:2007}, however, suggest
that the unresolved fluctuations in those simulations do not
contribute substantially to the stresses. In this case, the heating
prescription alone may be sufficient to determine the fate of the turbulent
energy lost numerically.

Turbulence in the interstellar medium can be probed by interpreting
the electron density fluctuations as a passive tracer mixed by the
ambient turbulence.  Interstellar scintillation measurements of
these density fluctuations in the diffuse ISM demonstrate an outer
scale turbulent wavenumber $k_0 \sim 1/L_0 \sim 10^{-15}$~km
\citep{Armstrong:1995}. Although the ISM is an inhomogeneous plasma
characterized by several different phases, it is the warm ionized
component of the ISM that responds to the turbulent electromagnetic
fluctuations. Taking typical plasma parameters of $n_i=n_e \sim
0.5$~cm$^{-3}$, $T_i\sim T_e \sim 8000$~K, and $B \sim 2 \times
10^{-6}$~G \citep{Schekochihin:2009}, we find $\rho_i \sim 500$~km and
$\beta_i \sim 3$. For $\beta_i =3$ and $T_i/T_e =1$, the driving
wavenumber of $k_0\rho_i \sim 10{^{-12}}$ easily satisfies the constraint in
Fig.~\ref{fig:k0}. Applying eq.~(\ref{eq:fit}), we predict an
ion-to-electron heating ratio of $Q_i/Q_e \simeq 1$ due to the
dissipation of turbulence in the warm ionized component of the ISM. As
seen in Fig.~\ref{fig:fit}, the predicted heating ratio is a
relatively strong function of $\beta_i$, so for any particular volume
of plasma, the heating may vary as $\beta_i$ varies, but it is clear
from the plot that in the range around $\beta_i \sim 3$ neither the
ions nor the electrons are dominantly heated. Although there is a
substantial cold, weakly ionized component of the ISM, collisional
coupling with the warm ions is extremely weak at the dissipative scales, so
it is unlikely to affect this prediction significantly.

\section{Conclusion}
This Letter presents a calculation of the ratio of ion to electron
heating $Q_i/Q_e$ due to the kinetic dissipation of Alfv\'enic
turbulence in astrophysical plasmas based on the turbulent cascade
model of \citet{Howes:2008b}. The results demonstrate that the heating
ratio $Q_i/Q_e$ is primarily a monotonic function of $\beta_i$ and has
weak dependence on $T_i/T_e$, as seen in Fig.~\ref{fig:qiqe}. 
A fit of the cascade model calculations is given by
eq.~(\ref{eq:fit}), a form suitable for use as a sub-grid-scale model
for turbulent heating in simulations of astrophysical plasmas,
potentially enabling the determination of observational signatures
caused by radiation emitted from the turbulently heated plasma.  Future work aims to employ nonlinear kinetic simulations of the
dissipation of astrophysical turbulence to test these predictions of the
turbulent plasma heating.




\label{lastpage}

\end{document}